\documentclass[aps,prc,twocolumn,groupedaddress]{revtex4-1}

\usepackage{graphicx}
\usepackage{amsmath}

\begin{document}


\title{Irregularities in nuclear radii at magic numbers}


\author{H. Nakada}
\email[E-mail:\,\,]{nakada@faculty.chiba-u.jp}
\affiliation{Department of Physics, Graduate School of Science,
 Chiba University,\\
Yayoi-cho 1-33, Inage, Chiba 263-8522, Japan}


\date{\today}

\begin{abstract}
Influence of magic numbers on nuclear radii is investigated
via the Hartree-Fock-Bogolyubov calculations and available experimental data.
With the $\ell s$ potential including additional density-dependence
suggested from the chiral effective field theory,
kinks are universally predicted at the $jj$-closed magic numbers
and anti-kinks (\textit{i.e.} inverted kinks) are newly predicted
at the $\ell s$-closed magic numbers,
both in the charge radii and in the matter radii
along the isotopic and isotonic chains where nuclei stay spherical.
These results seem consistent with the kinks of the charge radii
observed in Ca, Sn and Pb
and the anti-kink in Ca.
The kinks and the anti-kinks could be a peculiar indicator for magic numbers,
discriminating $jj$-closure and $\ell s$-closure.
\end{abstract}


\maketitle



\section{Introduction\label{sec:intro}}

As finite quantum many-body systems,
atomic nuclei show notable irregularities in their properties.
The most typical and significant example is magic numbers,
which have been identified in irregularities
of masses and excitation energies~\cite{ref:Hey04}.
The magic numbers are fundamental to nuclear structure physics.
Furthermore, they are relevant to the origin and abundance of elements,
forming waiting points in several processes of nucleosynthesis
in the universe~\cite{ref:syn97}.
Proton magicity in large neutron excess plays a key role
in the structure of the neutron-star crust as well~\cite{ref:NV73,ref:ODC08}.
Developments of the radioactive nuclear beams in recent decades
have disclosed that magic numbers are not so rigorous as once expected.
While some of the known magic numbers disappear,
new magic numbers come out far off the stability~\cite{ref:SP08}.
It is highly desired to comprehend magic numbers
all over the nuclear chart.

Though interesting and significant,
quantum irregularities as arising at the magic numbers are often an obstacle
to constructing an accurate and practical theoretical model.
One of the successful methods to handle quantum many-body problems
is the density functional theory,
as well-developed for bound electronic systems
in attractive external fields~\cite{ref:ED11}.
However, whereas the Hohenberg-Kohn theorem guarantees
the existence of an energy density functional (EDF)
that gives exact ground-state energies~\cite{ref:HK64},
it is crucial to remove irregularities properly,
as done by the Kohn-Sham method for the electronic systems~\cite{ref:KS65}.
Although there have been many attempts to construct nuclear EDFs
in terms of the nucleonic densities and quasi-local currents
(\textit{e.g.} Ref.~\cite{ref:UNEDF2}),
none have been as successful as in the electronic systems.
If a nuclear EDF could be developed
that attains accuracy at a comparable level to the electronic EDF,
it would provide us with a unified theoretical framework
for many-fermion systems.
Full understanding of irregularities such as magic numbers
is crucial also in this respect.

Nuclear radii are basic physical quantities,
directly linked to the density distributions
that are essential ingredients of an EDF.
The measured matter radii of stable nuclei are proportional
to the one-third of the mass number $A$ in the first approximation,
which manifests the saturation of the nucleon-number densities~\cite{ref:Hey04}.
Deviation from this simple rule carries interesting information
concerning nuclear structure.
For instance, some of the nuclei in the vicinity of the neutron drip line
have significantly large root-mean-square (rms) matter radii,
indicating neutron halos~\cite{ref:halo}.
Although the nuclear Hamiltonian should keep the rotational invariance,
a number of nuclei are deformed rather than spherical
in their intrinsic states.
The nuclear deformation has been verified by the distinctly large radii
compared to those of the nearby spherical nuclei,
together with other observables~\cite{ref:BM1}.

Since nuclei with magic proton ($Z$) or neutron numbers ($N$)
are usually spherical
and those without magicity often depart from the sphericity,
it is not surprising that the radii become relatively small
at magic $Z$ or $N$.
In practice, kinks have been observed at magic $N$
in the charge radii in many isotopes~\cite{ref:Ang13}.
However, kinks have been found at magic numbers
even when nuclei stay spherical.
A well-known example is a kink at $N=126$
in the isotope shifts of Pb~\cite{ref:AHS87}.
As discussed below, the relevance of the magicity to the nuclear radii
has not been elucidated sufficiently.
It is significant to perceive the presence and mechanism of irregularities
in the radii at magicity.

In this article, I shall discuss
the relationship between the radii of spherical nuclei and the magicity,
emphasizing possible roles of the three-nucleon ($3N$) interaction.
Originating from the nucleonic interaction,
nuclear EDFs are often associated with effective interactions.
A valuable guide for nuclear EDFs will be obtained
by appreciating how the nucleonic interaction affects nuclear structure.
Based on the predictions in which the $3N$-force effects are taken account of,
irregularities in radii are proposed as an experimental tool
that is useful for recognizing characters of individual magicity,
as well as for identifying magic numbers.
Note that magic numbers can well be identified by no single observable,
and the consistency among relevant physical quantities
should be checked carefully.
Whereas the magic numbers are usually identified via energies,
irregularities in radii could be important as well,
both experimentally and theoretically.

\section{Effects of spin-orbit potential on nuclear radii\label{sec:lspot}}

The spin-orbit ($\ell s$) splitting of the nucleon orbitals
is essential to the magic numbers.
Although it must be traced back to the nucleonic interaction,
the origin of the $\ell s$ splitting has not been understood
sufficiently~\cite{ref:AB81}.
While many-body correlations induced by the two-nucleon ($2N$) interaction were
suggested to enhance the $\ell s$ splitting~\cite{ref:LS-UMOA},
there was an argument from the quark-meson coupling model
that the $\rho$-meson exchange could contribute~\cite{ref:GT04}.
Via an \textit{ab initio} calculation applying the quantum Monte Carlo method,
the importance of the $3N$ interaction in the $\ell s$ splitting
was indicated~\cite{ref:LS-MC}.
It was argued recently,
based on the chiral effective-field theory ($\chi$EFT),
that the $3N$ interaction may account for
the missing part of the $\ell s$ splitting~\cite{ref:Koh13}.
The $\ell s$ splitting is linked to the $\ell s$ potential
at the mean-field (MF) level.

It has been recognized from the experimental data
that kinks often come out in the neutron-number ($N$) dependence
of the nuclear charge radii at $N=28,50,82$~\cite{ref:Ang13}
and likely at $N=126$~\cite{ref:BFI18},
while not so at $N=20$.
It should be noticed that the numbers $28,50,82$ and $126$ are
the $jj$-closed magic numbers,
in which a $j=\ell+1/2$ orbit is fully occupied
while its $\ell s$ partner is empty.
One of the well-known examples of the kinks
is found at $N=126$ in the Pb isotopes
[see Fig.~\ref{fig:drc_Z-magic}(d) below].
Since deformation is unlikely around $^{208}$Pb,
the kink implies that the proton wave-functions
are influenced by neutrons which occupy the single-particle (s.p.) levels
above the $N=126$ shell gap.
However, not all the neutron s.p. orbitals above the shell gap distribute
so broadly as to account for the observed kink.
It has been pointed out that neutron occupancy on the $0i_{11/2}$ orbit
is relevant to the kink,
whereas the s.p. function of $n1g_{9/2}$,
the lowest s.p. level above $N=126$~\cite{ref:TI}, is not broad enough
in the self-consistent MF calculations~\cite{ref:TBP93,ref:GSR13}.
The spatial distributions of the s.p. orbits
are affected by the $\ell s$ potential to a certain extent.
The $\ell s$ potential is repulsive (attractive) for a nucleon occupying
a $j=\ell-1/2$ ($j=\ell+1/2$) orbital,
tending to shift the wave function outward (inward).
This effect is the stronger for the higher $\ell$.
Thereby occupation of a $j=\ell-1/2$ orbit (\textit{e.g.} $0i_{11/2}$)
yields a larger radius than the occupation of surrounding orbitals.
Sizable occupation on $n0i_{11/2}$ in $N>126$ broadens the neutron wave-function
and may induce a rapid rise of the charge radii
through the attraction between protons and neutrons.

For the occupation on $n0i_{11/2}$ above $N=126$,
the s.p. energy difference relative to $n1g_{9/2}$ is crucial.
In the calculations with the conventional Skyrme EDFs,
$n0i_{11/2}$ lies significantly higher than $n1g_{9/2}$,
preventing the visible kink from emerging.
Via comparison to the relativistic mean-field (RMF) calculations
which yield a kink at $^{208}$Pb~\cite{ref:SLR94},
it was found that the $n0i_{11/2}$ occupation is related
to the isospin partitions of the $\ell s$ potential~\cite{ref:SLKR95,ref:RF95},
which should originate from a certain channel of the nucleonic interaction.
Still, it has been difficult to reproduce the kink
unless $n1g_{9/2}$ and $n0i_{11/2}$ are
nearly degenerate or even inverted~\cite{ref:GSR13},
incompatible with the observed energy levels~\cite{ref:TI}.
On the contrary, if there is a significant contribution of the $3N$ interaction
to the $\ell s$ potential as suggested by the $\chi$EFT,
it makes the $\ell s$ potential stronger in the nuclear interior
than in the exterior.
Then the difference in the radial distribution between the $\ell s$ partners
is grown further,
because the wave-functions of the $j=\ell-1/2$ ($j=\ell+1/2$) orbitals
slide outward (inward) so as to be influenced by the $\ell s$ potential
to a lesser (greater) extent.
This effect on the s.p. functions has been confirmed
in Fig.~1 of Ref.~\cite{ref:NI15}.
The enhanced difference of the wave-functions improves
$N$-dependence of the charge radii in Pb
with little influence on the s.p. energies~\cite{ref:NI15}.
Similarly, the kink of the charge radii in Ca at $N=28$ is pronounced
and a kink is predicted in Sn at $N=82$~\cite{ref:Nak15}
[see also Fig.~\ref{fig:drc_Z-magic}(a) and (c) below].
Both kinks have been observed
in recent experiments~\cite{ref:shift-Ca52,ref:shift-Sn134}.
The kink at $^{48}$Ca has been obtained by \textit{ab initio} methods
with the $\chi$EFT interactions as well~\cite{ref:shift-Ca52}.

There are two types of nuclear magic numbers:
the $\ell s$-closed magic numbers and the $jj$-closed ones.
While magicity is normally indicated by irregularities in energies
that do not discern between the $\ell s$-closed
and the $jj$-closed magic numbers,
the irregularities in the nuclear radii may work as a peculiar indicator.
The $jj$-closed magic numbers occur after a high-$j$ orbit with $j=\ell+1/2$
is fully occupied,
and its $\ell s$ partner with $j=\ell-1/2$ starts occupied
above the magic numbers.
Even though the $j=\ell-1/2$ orbit does not always lie lowest
above the magic number,
its occupancy is sizable owing to the pair correlation.
Therefore the nuclear radii increase relatively slowly below the magicity
and more rapidly above it, producing a kink.
On the other hand, the $\ell s$-closed magic numbers occur
after a $j=\ell-1/2$ orbit is filled,
and a $j=\ell+1/2$ orbit with higher $\ell$ starts being occupied above it.
It is then expected that the nuclear radii increase rapidly
below the $\ell s$-closed magic numbers,
and increase more slowly or even decrease above it.
Thus an inverted kink emerges at the $\ell s$-closed magic numbers,
which will be called `anti-kink' in contrast to the kink
at the $jj$-closed magic numbers.
As well as the magicity itself,
its character, \textit{i.e.} whether it is $jj$- or $\ell s$-closed,
may be examined by qualitative behavior of the nuclear radii.
The kinks and the anti-kinks are expected to be visible or pronounced
as an effect of the $3N$ interaction,
as shown in Sec.~\ref{sec:result}.

While accurate data can be obtained for the charge radii,
experimental data on the matter radii have been reported
for some isotopic chains (\textit{e.g.} Ref.~\cite{ref:OST01}).
More abundant data including unstable nuclei
are expected in future experiments using hadronic probes.
Nuclear matter radii are an average
reflecting the radial distributions of all the constituent nucleons.
It is also intriguing whether and how the neutron magicity influences
isotopic variation of the nuclear matter radii,
which directly reflect the radial distributions of neutrons.
The same holds for the proton magicity under the isotonic variation.

It should be kept in mind that
deformation can be another source of irregularities in the nuclear radii.
As the deformation is suppressed at the magic numbers,
it tends to produce a kink, not an anti-kink.
For the $\ell s$-closed magicity,
the effects of the s.p. functions and the deformation may act competitively,
possibly obscuring the anti-kinks.
Halos, which could emerge in vicinity of the drip lines,
also give rise to irregularity in nuclear radii.
However, it will be feasible to investigate the magicity via the radii,
by choosing a series of spherical nuclei not too close to the drip lines.

\section{Theoretical and experimental results\label{sec:result}}

Let us see how the above arguments apply
to the theoretical and experimental results.
To illustrate kinks and anti-kinks theoretically,
I shall present results of self-consistent MF calculations,
the spherical Hartree-Fock-Bogolyubov (HFB) to be precise~\cite{ref:Nak06},
for nuclei having magic $Z$ or $N$.
Odd-$A$ nuclei are treated
in the equal-filling approximation~\cite{ref:EFA,ref:EFA2}.
For the nucleonic effective interaction,
the M3Y-P6 and M3Y-P6a semi-realistic interactions~\cite{ref:Nak13,ref:NI15}
are mainly employed.
For comparison, results with the Gogny-D1S interaction~\cite{ref:D1S},
which has been one of the most widely-used interactions
for the HFB calculations,
are also displayed.
Influence of the center-of-mass motion is corrected~\cite{ref:NS02}.
For the charge radii,
the finite-size effects of the constituent nucleons are taken into account,
up to the magnetic effects~\cite{ref:FN75}.
Also for reference, results of the RMF calculations
for even-even nuclei with the NL3 parameter
are quoted from Ref.~\cite{ref:NL3},
in which some of the finite-size effects on the charge radii are ignored.

In the self-consistent MF framework,
the $\ell s$ splitting is obtained primarily from the LS channel
of the nucleonic interaction,
\begin{equation}
v_{ij}^{(\mathrm{LS})} = \sum_n \big(t_n^{(\mathrm{LSE})} P_\mathrm{TE}
 + t_n^{(\mathrm{LSO})} P_\mathrm{TO}\big) f_n^{(\mathrm{LS})} (r_{ij})\,
 \mathbf{L}_{ij}\cdot(\mathbf{s}_i+\mathbf{s}_j)\,, \label{eq:LS}
\end{equation}
within the $2N$ interaction.
Here the subscripts $i$ and $j$ are indices of nucleons.
$P_\mathrm{TE}$ ($P_\mathrm{TO}$) denotes the projection operator
on the triplet-even (triplet-odd) two-particle states,
$\mathbf{r}_{ij}= \mathbf{r}_i - \mathbf{r}_j$, $r_{ij}=|\mathbf{r}_{ij}|$,
$\mathbf{p}_{ij}= (\mathbf{p}_i - \mathbf{p}_j)/2$,
$\mathbf{L}_{ij}= \mathbf{r}_{ij}\times \mathbf{p}_{ij}$,
and $\mathbf{s}_i$ is the spin operator.
In the M3Y-type interactions,
$f_n^{(\mathrm{LS})}(r)=e^{-\mu_n^{(\mathrm{LS})} r}/\mu_n^{(\mathrm{LS})} r$,
with $1/\mu_n^{(\mathrm{LS})}$ representing the range parameter~\cite{ref:Nak03}.
In M3Y-P6, which gives a reasonable prediction of magic numbers
in a wide range of the nuclear chart
including unstable nuclei~\cite{ref:NS14},
the strength parameters $t_n^{(\mathrm{LSE})}$ and $t_n^{(\mathrm{LSO})}$
derived from Paris $2N$ force~\cite{ref:M3Y-P}
are multiplied by a factor $2.2$,
to reproduce the level sequence around $^{208}$Pb.
On the other hand, analysis based on the $\chi$EFT suggests
that the $3N$ interaction enhances the LS channel
so that it should become stronger
as the nucleon density increases~\cite{ref:Koh13}.
Hinted by this result, in M3Y-P6a
a density-dependent term $v^{(\mathrm{LS}\rho)}$ is added
instead of enhancing $t_n^{(\mathrm{LSE})}$ and $t_n^{(\mathrm{LSO})}$,
which is represented as~\cite{LSrho_correction}
\begin{equation}
\begin{split}
v_{ij}^{(\mathrm{LS}\rho)} =& 2i\,D[\rho(\mathbf{R}_{ij})]\,
 \mathbf{p}_{ij}\times\delta(\mathbf{r}_{ij})\,\mathbf{p}_{ij}\cdot
 (\mathbf{s}_i+\mathbf{s}_j)\,;\\
& D[\rho(\mathbf{r})] = w_1\,\frac{\rho(\mathbf{r})}
{1+d_1\rho(\mathbf{r})}\,.
\end{split}\label{eq:DDLS}
\end{equation}
Here $\rho(\mathbf{r})$ is the isoscalar nucleon density
and $\mathbf{R}_{ij}=(\mathbf{r}_i+\mathbf{r}_j)/2$.
The density-dependent coefficient $D[\rho]$
carries effects of the $3N$ interaction.
The parameter $w_1$ is fitted to the $n0i_{13/2}$-$n0i_{11/2}$ splitting
with M3Y-P6 at $^{208}$Pb.
Then the s.p. energies, as well as the binding energies, do not change
from those of M3Y-P6 significantly.
The parameter $d_1$ does not have physical importance,
and $d_1=1.0\,\mathrm{fm}^3$ is assumed~\cite{ref:Nak15}.
As all the channels except the LS one are identical
between M3Y-P6 and M3Y-P6a,
comparison of their results will clarify effects of the $3N$ LS term
[\textit{i.e.} $v^{(\mathrm{LS}\rho)}$]
in place of the naive enhancement of the LS channel by an overall factor.
While the form of Eq.~(\ref{eq:DDLS}) is consistent
with the $\chi$EFT analysis~\cite{ref:Koh13}
by which the qualitative effects of the $3N$ interaction could be investigated,
the strength is not equal to that derived in Ref.~\cite{ref:Koh13}.
Both M3Y-P6 and M3Y-P6a contain realistic tensor channels
based on the $G$-matrix~\cite{ref:M3Y-P},
which also have influence on the $\ell s$ splitting
in some nuclei~\cite{ref:NS14,ref:Vtn,ref:SC14,ref:Nak10,ref:NSM13}.

\begin{figure}
\includegraphics[scale=0.25]{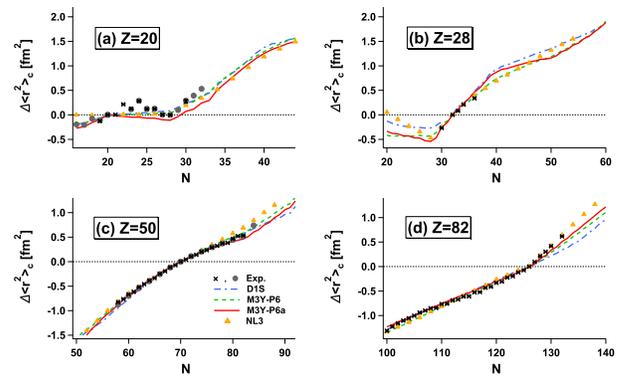}
\caption{(Color online) $N$-dependence of differential mean-square charge radii
  of (a) $Z=20$, (b) $Z=28$, (c) $Z=50$ and (d) $Z=82$ nuclei.
  Spherical HFB results with D1S (blue dot-dashed line),
  M3Y-P6 (green dashed line) and M3Y-P6a (red solid line) are presented.
  RMF results for even-$N$ nuclei are quoted
  from Ref.~\cite{ref:NL3} (orange triangles).
  Experimental data are taken from Refs.~\cite{ref:Ang13} (crosses)
  for all panels,
  \cite{ref:shift-Ca36} (for $N\leq 20$)
  and \cite{ref:shift-Ca52} (for $N\geq 23$) in (a) (gray circles),
  \cite{ref:shift-Sn134} in (c) (gray circles).
\label{fig:drc_Z-magic}}
\end{figure}

The nuclear charge radii can be measured by the electromagnetic probes,
\textit{e.g.} the electron scattering~\cite{ref:FN75}.
Moreover, the mean-square differential charge radii among isotopes,
denoted by $\mathit{\Delta}\langle r^2\rangle_c$,
are extracted accurately from the isotope shifts~\cite{ref:Ang13}.
For the nuclide $^A Z$,
$\mathit{\Delta}\langle r^2\rangle_c$ is defined by
$\mathit{\Delta}\langle r^2\rangle_c(^A Z):=
\langle r^2\rangle_c(^A Z)-\langle r^2\rangle_c(^{A_0} Z)$,
where $^{A_0} Z$ is the reference nuclide for the fixed $Z$.
In Fig.~\ref{fig:drc_Z-magic},
$\mathit{\Delta}\langle r^2\rangle_c$ in the magic-$Z$ nuclei
are plotted as a function of $N$.
As reference nuclei,
$^{40}$Ca, $^{60}$Ni, $^{120}$Sn and $^{208}$Pb are taken
as in Refs.~\cite{ref:Ang13}.
Experimentally, kinks have been observed at $^{48}$Ca, $^{132}$Sn and $^{208}$Pb
as already mentioned,
corresponding to the neutron $jj$-closed magic numbers.
In the theoretical results,
interaction-dependence is found for the kinks.
In Pb, the isospin-dependence of the $\ell s$ potential
affects $\mathit{\Delta}\langle r^2\rangle_c$
around $N=126$~\cite{ref:SLKR95}
through the s.p. energy difference
$\varepsilon_n(0i_{11/2})-\varepsilon_n(1g_{9/2})$.
The D1S interaction has the zero-range LS channel
as the Skyrme interaction~\cite{ref:Sky},
yielding no apparent kink at $N=126$ in Fig.~\ref{fig:drc_Z-magic}(d).
A kink is obtained at $N=126$ with M3Y-P6,
but it is weaker than the observed one.
The kink becomes pronounced in the M3Y-P6a results~\cite{ref:NI15}
without significant change in $\varepsilon_n(0i_{11/2})-\varepsilon_n(1g_{9/2})$
from M3Y-P6,
which is comparable to the observed energy difference
between $(11/2)^+$ and $(9/2)^+$ at $^{209}$Pb~\cite{ref:TI}.
Kinks universally arise at the $jj$-closed magicity with M3Y-P6a,
\textit{i.e.} by taking into account the $3N$-force contribution
to the LS channel that affects the s.p. functions.
Note that this is not the case for the RMF results of Ref.~\cite{ref:NL3}.
As pointed out in Ref.~\cite{ref:Nak15},
a kink has been predicted at $N=82$ for the Sn chain with M3Y-P6a
[Fig.~\ref{fig:drc_Z-magic}(c)],
though such a prominent kink is not seen in the other results shown here.
The recent discovery of a kink at $^{132}$Sn~\cite{ref:shift-Sn134}
is supportive of the $3N$-force contribution to the $\ell s$ splitting.
A kink is also predicted at $N=28$ for the Ni chain
[Fig.~\ref{fig:drc_Z-magic}(d)],
which is generic for interactions but enhanced
by introducing $v^{(\mathrm{LS}\rho)}$.
Moreover, anti-kinks are grown at $^{40}$Ca and $^{68}$Ni
in the M3Y-P6a results,
because of the $\ell s$-closed magicity of $N=20$ and $40$.
The former is indeed consistent
with the recent measurement~\cite{ref:shift-Ca36}
as exhibited in Fig.~\ref{fig:drc_Z-magic}(a).
The anti-kinks are of particular importance in establishing
effects of the magicity on the nuclear radii
and roles of the $3N$ interaction in them.
For the $N=40$ magicity,
no obvious anti-kink is seen at $^{60}$Ca even with M3Y-P6a,
since the magicity is not well kept at $^{60}$Ca~\cite{ref:NS14,ref:Nak10}.
The kink-like structure at $^{54}$Ca might be related
to the $N=34$ magicity~\cite{ref:Ca54_Ex2},
although it was not identified as magic in Ref.~\cite{ref:NS14}.

\begin{figure}
\includegraphics[scale=0.25]{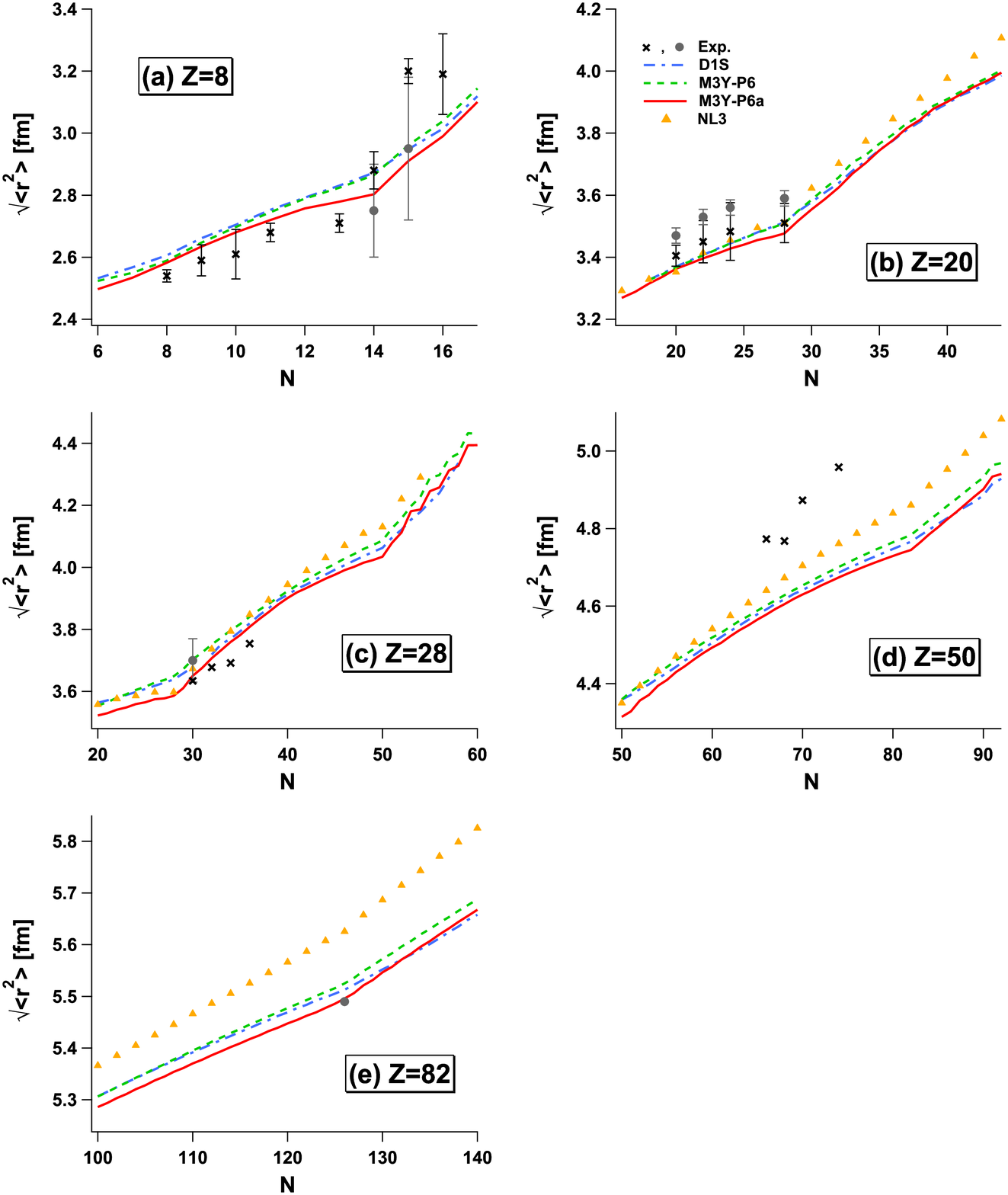}
\caption{(Color online) $N$-dependence of rms matter radii
  of (a) $Z=8$, (b) $Z=20$, (c) $Z=28$, (d) $Z=50$
  and (e) $Z=82$ nuclei.
  For conventions for theoretical results, see Fig.~\ref{fig:drc_Z-magic}.
  Experimental data are taken from
  Refs.~\cite{ref:OST01} (crosses) and \cite{ref:Kan11} (gray circles) in (a),
  \cite{ref:MNB86} (crosses) and \cite{ref:Alk77} (gray circles) in (b),
  \cite{ref:CLS77} (crosses) and \cite{ref:Zam17} (gray circles) in (c),
  \cite{ref:BFW71} (crosses) in (d),
  and \cite{ref:ABV78} (gray circles) in (e).
\label{fig:r_Z-magic}}
\end{figure}

\begin{figure}
\includegraphics[scale=0.25]{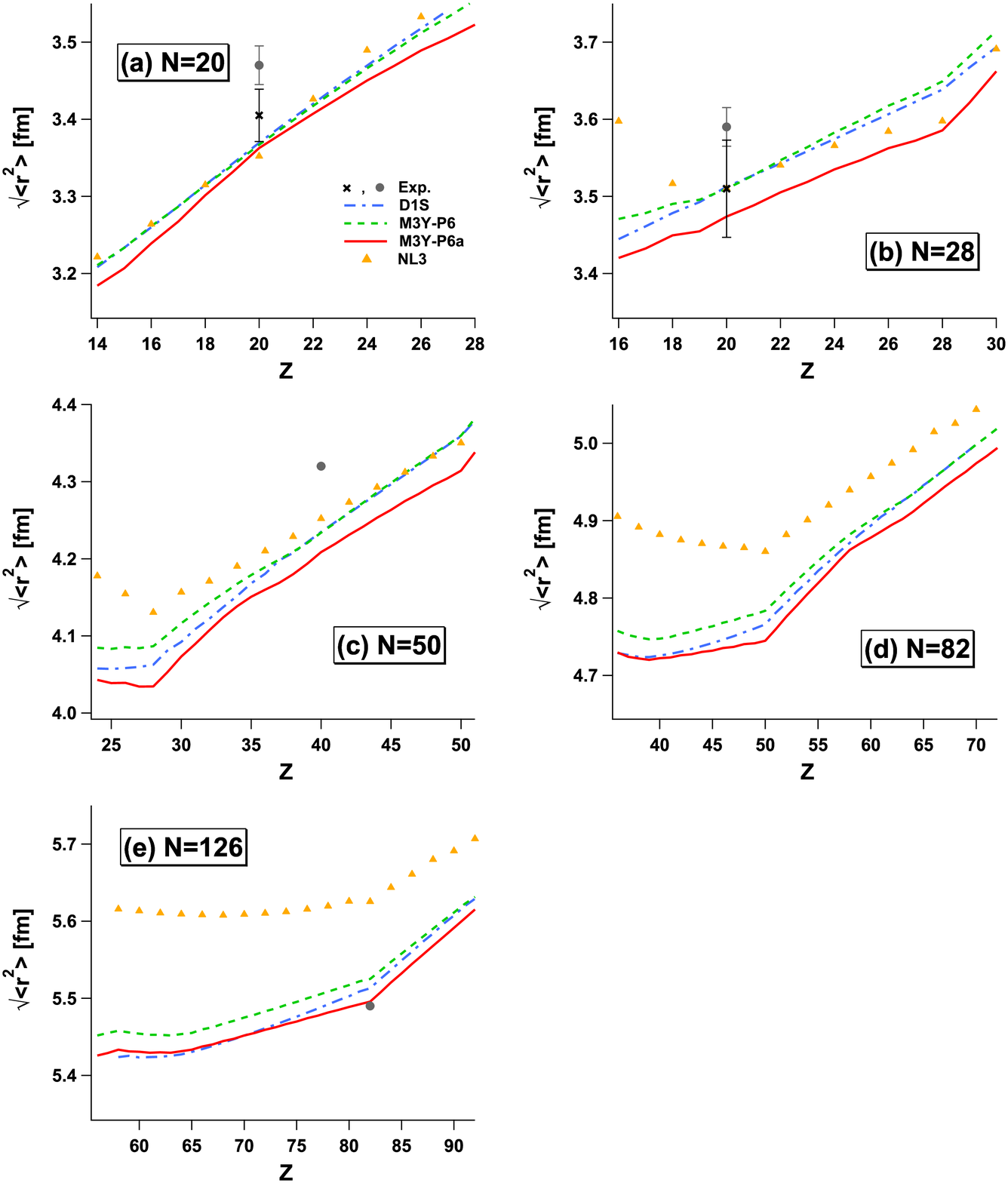}
\caption{(Color online) $Z$-dependence of rms matter radii
  of (a) $N=20$, (b) $N=28$, (c) $N=50$, (e) $N=82$
  and (e) $N=126$ nuclei.
  For conventions for theoretical results, see Fig.~\ref{fig:drc_Z-magic}.
  Experimental data are taken from
  Ref.~\cite{ref:MNB86} (crosses) and \cite{ref:Alk77} (gray circles) in (a,b),
  and \cite{ref:ABV78} (gray circles) in (c,e).
\label{fig:r_N-magic}}
\end{figure}

$N$-dependence ($Z$-dependence) of the rms matter radii
is depicted for the magic-$Z$ (magic-$N$) nuclei
in Fig.~\ref{fig:r_Z-magic} (Fig.~\ref{fig:r_N-magic}).
Not so many data are available for the matter radii,
and it has not been easy to attain good accuracy.
However, owing to the progress in experimental techniques and reaction theory,
systematic measurements with good precision are promising,
up to nuclei far off the $\beta$-stability. 
Future experiments over isotopic or isotonic chains are awaited.

In Fig.~\ref{fig:r_Z-magic}(a),
a kink is predicted at $N=14$, which is enhanced by $v^{(\mathrm{LS}\rho)}$.
This corresponds to the submagic nature of $N=14$ at $^{22}$O~\cite{ref:NS14}.
Although this kink seems compatible
with the available data~\cite{ref:OST01,ref:Kan11},
more accurate data are desirable.
For the other isotopic chains,
kinks are predicted at the usual $jj$-closed magic numbers.
While the kinks are weak without $v^{(\mathrm{LS}\rho)}$,
they come pronounced in the M3Y-P6a results.
It is mentioned that the kink at $^{48}$Ca is observed
in a recent experiment~\cite{TF-pv}.
Anti-kinks are predicted with M3Y-P6a
at $N=20$ in Fig.~\ref{fig:r_Z-magic}(b)
and at $N=40$ in Fig.~\ref{fig:r_Z-magic}(c),
corresponding to the $\ell s$ closure.
Not apparent in the other results,
the anti-kinks can disclose the $3N$-force effects,
although these anti-kinks are less conspicuous
than the kinks at the $jj$-closed magicity.

In Fig.~\ref{fig:r_N-magic}(a),
an anti-kink is predicted with M3Y-P6a at $Z=20$,
linked to the $\ell s$-closed magicity.
No visible anti-kink is predicted at $^{48}$Ca in Fig.~\ref{fig:r_N-magic}(b).
This is accounted for by the inversion of the s.p. levels
$p0d_{3/2}$ and $p1s_{1/2}$~\cite{ref:NSM13}.
In Fig.~\ref{fig:r_N-magic}(c),
a kink is predicted at the $jj$-closed magic number $Z=28$.
Figure~\ref{fig:r_N-magic}(d) shows several irregularities.
In addition to a kink at the $Z=50$ magicity,
an anti-kink is predicted at $Z=58$
and a weak kink is viewed at $Z=64$.
The former corresponds to the closure up to $p0g_{7/2}$ at $^{140}$Ce
and the latter to the closure of $p1d_{5/2}$ at $^{146}$Gd,
both of which are identified as submagic numbers in Ref.~\cite{ref:NS14},
consistent with the relatively high excitation energies
in measurement~\cite{ref:TI}.
The irregularities in the radii will support their submagic nature
if observed.
In Fig.~\ref{fig:r_N-magic}(e),
a kink is predicted at $Z=82$, irrespective of the interactions.
An anti-kink predicted at $Z=58$ and a weak kink at $Z=64$
are attributed to the submagic nature~\cite{ref:NS14},
as in the $N=82$ case.

It is commented that the kinks
in $\mathit{\Delta}\langle r^2\rangle_c$
have also been predicted with the Fayans EDF
at $^{48}$Ca, $^{132}$Sn and $^{208}$Pb~\cite{ref:Fay98}.
The Fayans EDF and M3Y-P6a are the only two EDFs
that have predicted the kink at $^{132}$Sn.
The results of $\mathit{\Delta}\langle r^2\rangle_c$
have similarity to those of M3Y-P6a in qualitative respect,
despite difference in the EDF forms.
Whereas the relation of the kinks to the nucleonic interaction is not clear
in the Fayans EDF,
effects of the pairing channel have been stressed~\cite{ref:shift-Sn134}.
This is not necessarily contradictory to the present analysis,
as the pairing plays a role in the occupation of the relevant s.p. orbits.
It is of interest whether the other results with M3Y-P6a shown here
are shared with those with the Fayans EDF.
However, the pairing should not strengthen the anti-kinks.
Future experiments around the $\ell s$-closed magicity
will be significant to pin down the dominant source of the irregularities.
Apart from the irregularities,
the RMF results of the matter radii
in Figs.~\ref{fig:r_Z-magic} and \ref{fig:r_N-magic}
are considerably larger than the others in the neutron excess,
whereas the charge radii are comparable.
These results indicate thick neutron skins with the NL3 parameter-set in the RMF
and are attributed to the strong density-dependence of the symmetry energy.

\section{Summary\label{sec:summary}}

Influence of magic numbers on nuclear radii has been investigated
via the self-consistent spherical Hartree-Fock-Bogolyubov (HFB) calculations
and available experimental data.
Owing to the difference in the single-particle wave-functions
between $\ell s$ partners,
kinks are universally expected at the $jj$-closed magic numbers
both in the charge radii and the matter radii,
even when nuclei stay spherical.
Although the former has been recognized empirically,
most of the HFB calculations do not reproduce all the kinks
at $N=28$, $82$ and $126$ in the Ca, Sn and Pb isotopes.
The density-dependence of the $\ell s$ potential,
which can be linked to the $3N$ interaction
suggested from the chiral effective field theory,
yields significant contribution to the kinks
which are consistent with the data.
Moreover, the calculations with this density-dependence
predict `anti-kinks' at the $\ell s$-closed magic numbers,
\textit{i.e.} kinks inverted from the $jj$-closed cases.
If experimentally established,
the anti-kinks could be good evidence
for the $3N$-force effects on the $\ell s$ splitting
and may be used to investigate nuclear magic numbers,
discriminating $jj$-closure and $\ell s$-closure
as well as indicating magicity.
Finally, it is stressed that appreciation
of effects of the magic numbers on irregularities in the radii,
such as the kinks and the anti-kinks,
is indispensable to construct an accurate and practical theory
using an energy density functional.

\begin{acknowledgments}
%
Discussions with M.~Fukuda, M.~Tanaka, S.~Shlomo and T.~Inakura
are gratefully acknowledged.
A part of numerical calculations is performed on HITAC SR24000
at IMIT in Chiba University.
\end{acknowledgments}


\begin{thebibliography}{99}
\bibitem{LSrho_correction}
  In Refs.~\protect\cite{ref:NI15,ref:Nak15},
  signs concerning $D[\rho]$ were wrong in several equations.
  The signs should be inverted in Eqs.(3,4) of Ref.~~\protect\cite{ref:NI15}
  and in Eqs.~(2,3) of Ref.~\cite{ref:Nak15}.
  The minus signs of Eqs.~(5) in both of Refs.~\protect\cite{ref:NI15,ref:Nak15}
  and that of $\delta D/\delta\rho$ just below it
  in Ref.~\protect\cite{ref:NI15} should be dropped.
\end{thebibliography}

\end{document}